\documentstyle[11pt,epsf]{article}
\newcommand{\etc}    {{\it etc.}}
\newcommand{\ie}     {i.\,e.}
\newcommand{\eg}     {e.\,g.}
\newcommand{\cf}     {{\it cf.\ }}
\newcommand{\araa}   {ARA\&A}
\newcommand{\aap}    {A\&A}
\newcommand{\apj}    {ApJ}
\newcommand{\apjl}   {ApJ(Letters)}

\newcommand{\mnras}  {MNRAS}

\newcommand{\um}     {\,$\mu\mbox{m}$}

\newcommand{\kms}    {\,$\mbox{km\,s}^{-1}$}
\newcommand{\nod}    {\,$\mbox{cm}^{-3}$}

\begin{document}

\noindent
{\small \em From Astrophysical Implications of the Laboratory Study of 
Presolar Materials \newline AIP Conference Proceedings, 1997
\newline T.J. Bernatowicz and E. Zinner (eds.)}

\begin{center}
{\huge \bf The Propagation and Survival\\ of Interstellar Grains}
\end{center}
 
\begin{center}
{\Large A.P. Jones$^{1,2}$, A.G.G.M. Tielens$^3$, D.J. Hollenbach$^3$\\ 
and C.F. McKee$^4$}
\end{center}

\begin{center}
{\small 
$^1$SETI Institute, 2035 Landings Drive, Mountain View, CA 94043, USA\\
$^2$IAS, Universit\'{e} Paris XI - B\^{a}timent 121, 91405 Orsay cedex, 
   France\\
$^3$NASA Ames Research Center, MS 245-3, Moffett Field, CA 94035, USA\\
$^4$Space Sciences Laboratory, University of California, Berkeley,
   CA 94720, USA}
\end{center}

\vspace*{0.5cm}

\begin{quote}
{\bf Abstract.}
In this paper we discuss the propagation of dust through the interstellar 
medium (ISM), and describe the destructive effects of stellar winds, jets, 
and supernova shock waves on interstellar dust. We review the probability
that grains formed in stellar outflows or supernovae survive processing 
in and propagation through the ISM, and incorporate themselves relatively
unprocessed into meteoritic bodies in the solar system.  We show that very 
large (radii $\geq 5$\um) and very small grains (radii $\leq 100$\AA) with 
sizes similar to the pre-solar SiC and diamond grains extracted from 
meteorites, can survive the passage through 100\kms\,shock waves 
relatively unscathed. High velocity ($\geq 250$\kms) shocks destroy dust 
efficiently. However, a small ($\sim 10$\%) fraction of the stardust never 
encountered such fast shocks before incorporation into the solar system. 
All grains should therefore retain traces of their passage through 
interstellar shocks during their propagation through the ISM. The grain 
surfaces should show evidence of processing due to sputtering and 
pitting due to small grain cratering collisions on the micron-sized 
grains. This conclusion seems to be in conflict with the evidence from 
the large grains recovered from meteorites which seem to show little 
interstellar processing. 
\end{quote}

\section{Introduction}

Dust formed in stellar environments (stardust and circumstellar dust) is 
primarily made in the shells around stars in the red giant phase of their 
evolution (\eg, M giants, carbon stars and radio luminous OH/IR stars), but 
some small fraction is also formed in the circumstellar shells around 
supergiants, novae, planetary nebulae (PN), WC stars, and in the ejecta 
of supernovae (SN types Ia and II). The presence of circumstellar dust 
is revealed by the large near IR continua arising from the heated dust 
in the vicinity of the star, and also by the opacity of the circumstellar 
shell which is due to the absorption and scattering of starlight by the 
local dust. In Figure 1 we show the relative contributions of the major 
sources of stardust in the ISM, and in Table 1 we give the stardust source 
contributions in units of $10^{-6}$ M$_{\odot}$ kpc$^{-2}$ yr$^{-1}$ [1,2].


\begin{figure}
 {\epsfxsize=\textwidth \epsffile{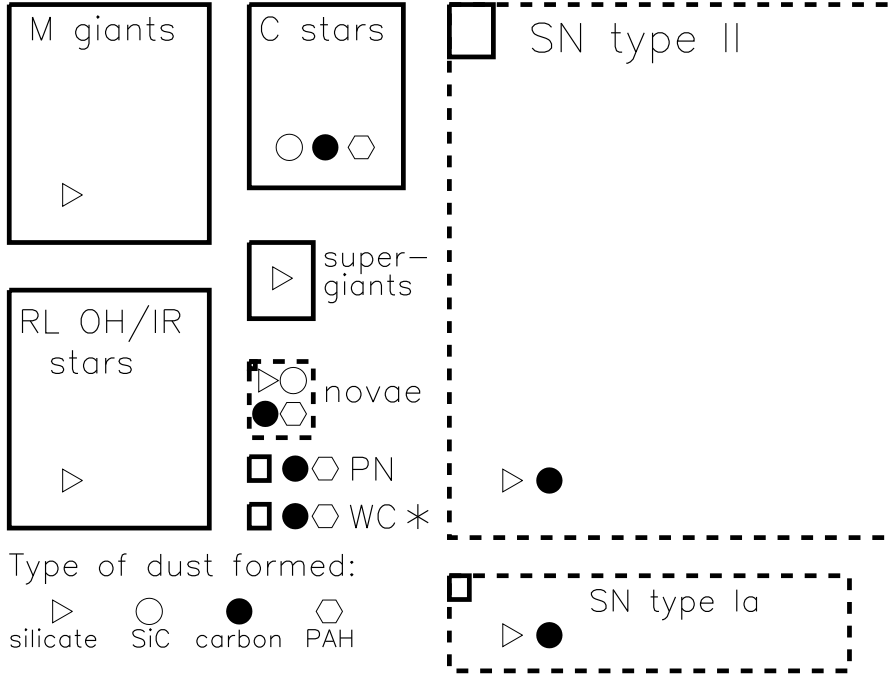}}
 \caption{Relative contributions, by boxed area, of stardust sources in 
 the ISM. C stars refers to carbon star sources, and PAH refers to 
 polycyclic aromatic hydrocarbon species. The dashed lines indicate 
 possible upper limit contributions.}
\end{figure}

%
\begin{table}
\caption{Contributions of Stardust Sources in the ISM}
\begin{center}
\begin{tabular}{lc} 
  \hline
  \multicolumn{1}{l}{Source}  & 
  \multicolumn{1}{c}
  {Contribution} \\
  \multicolumn{1}{l}{ }  & 
  \multicolumn{1}{c}
  {($\times 10^{-6}$ M$_{\odot}$ kpc$^{-2}$ yr$^{-1}$)} \\
  \hline
  M giants       &    3        \\ 
  RL OH/IR stars &    3        \\ 
  C stars        &    2        \\ 
  Supergiants    &   0.2       \\ 
  Novae          & 0.003$-$0.2 \\ 
  PN             &   0.03      \\ 
  WC stars       &   0.03      \\ 
  SN type II     & 0.15$-$14   \\ 
  SN type Ia     & 0.03$-$2.3  \\ 
  \hline
\end{tabular}
\end{center}
\end{table}

Stardust is injected into the ambient ISM by stellar winds. Observationally, 
stars with the highest mass loss rates show the presence of dust, implying 
a link  between high mass loss rates and the formation of dust. In essence, 
high mass loss rates correspond to high densities near the photosphere which 
are conducive to dust formation. Radiation pressure pushes the dust out, 
dragging the gas along, and this coupling is also favored by the high 
density [3].

In the ISM, dust is subject to processing in a variety of environments. 
This includes the growth of ice mantles, consisting of simple molecules 
such as H$_2$O, CO, CO$_2$, and CH$_3$OH, in the shielded environment of 
molecular clouds, and their subsequent processing into more complex organic 
mantles when exposed to FUV photons, and cosmic ray bombardment. Because 
of the low temperatures and densities, the dust formed in the ISM -- we 
shall call it ``ISM-dust" as opposed to stardust -- is formed far from 
equilibrium and will probably resemble kerogen, in terms of structure, 
presence of impurities, \etc, rather than graphite. Also, because the Si/C 
abundance ratio is much less than unity, no silicates or SiC grains are 
expected to form in the ISM (although silicon may well be incorporated into 
mantle materials [4]). Interstellar dust,  the dust which resides in the 
ISM, is then a mixture of ISM-dust and stardust.

Information on interstellar dust can be gleaned from observations of
stellar reddening, spectral absorption features, and the gas phase 
abundances of (rock--forming) elements [5]. In particular, the interstellar 
grain size distribution is commonly represented by a steep power law 
($n(a)\propto a^{-3.5}$ [6], MRN) which extends into the molecular domain 
(\ie, PAHs [7]). Because of the limited amount of dust-forming elements, 
and the fact that such a power law distribution has most of the mass in the 
largest particles, there must be a maximum grain size cut off beyond which 
the interstellar grain size distribution steepens considerably [8]. This 
maximum size is of order 3000\AA. There have been many reviews in recent 
years on the subject of interstellar dust [9--16] and the reader is 
referred to these, and the references therein, for further details. In this 
paper, we are mainly concerned with the destruction of dust in the diffuse ISM. 

We are most interested in the  dust in the warm intercloud component of the 
three-phase model of the ISM [17], because it is in this phase of the ISM 
that the bulk of the grain destruction  occurs [18--22]. The three-phase 
model of the ISM consists of a hot intercloud medium, a warm 
ionized/neutral medium (or ``warm intercloud medium"), and the cold
neutral medium.  For the hot, warm, and cold phases, respectively, the  
volume filling factors are  $\sim$ 0.6, $\sim$ 0.4, and $\sim$ 
0.02--0.04, the mean hydrogen atom densities, $n_{\rm H}$, are $\sim 0.003, 
\sim 0.25,$ and$ \sim 40$ \nod, and the mean gas kinetic temperatures, 
$T_{\rm k}$, are $\sim 5 \times 10^5, \sim 10^4,$ and $\sim 80$ K [17]. 
The cold medium rapidly cycles with the warm medium, on 
time scales of order $3 \times 10^7$ yr,  due to the photodissociation and 
photoionization of diffuse and molecular clouds by massive stars [21].  
The warm medium cycles back to the cold medium by recombination in
shielded regions and by shock compression. In a steady state, the warm 
medium, with about 10\% of the mass, must cycle to the cold medium on time 
scales of order $3 \times 10^6$ yr. The mass in the hot phase is too low for 
significant grain destruction to occur within the lifetime of a supernova 
remnant, and the small filling factor and low shock velocities of the cold 
component ensure little contribution to the grain destruction from this 
phase. It is therefore grain processing in the warm ISM, through the 
effects of supernova shock waves, that determines the lifetime of
grains in the ISM.

\section{Stardust and ISM-Dust in the Solar System}

In the ISM we have direct evidence for amorphous silicate grain materials
(both stardust and ISM-dust) from the Si--O bond stretching and
bending modes at 10\um\ and 20\um\ observed in absorption and emission
in dusty stellar envelopes, and in absorption toward stars without dusty
circumstellar envelopes. There is evidence for hydrocarbon grains from the 
3.4\um\ C--H stretching vibration which is seen in absorption along lines 
of sight through the diffuse ISM. The bump in the extinction curve at 
2175\AA\ is also attributed to C--bearing dust, generally of graphitic 
composition.  We also have indirect evidence for carbon-bearing dust from 
the depletion  of carbon in the diffuse ISM. It has been suggested [23]  
that the grains known as GEMS (Glass with Embedded Metal and Sulphides) 
found in the collected interplanetary dust particles represent silicate 
stardust transported through the ISM. This suggestion, whilst intriguing, 
is still not without problems. A detailed overview of the evidence for, 
and against, GEMS being interstellar material has been presented [24].

Stardust is clearly detected in the solar system. The evidence from 
meteorites shows that presolar grains were transported through the ISM and 
incorporated into the solar nebula $4.5$ billion years ago. Unequivocally, 
dust that formed in circumstellar environments and survived for some 
considerable time in the ISM has been recovered from meteorites [25] 
(also, see the extensive reviews elsewhere in these proceedings). 
In Table 2 we summarize the extracted and analyzed refractory 
presolar grain components of the Orgueil meteorite (data taken from 
[26] and references therein). The presolar grains described in Table 2 are, 
however, only trace components making up less than 0.15\% of the meteorite 
mass. Presumably much of the less refractory  ISM-dust and stardust 
(\eg, silicates) incorporated into the meteorite is lost during the 
chemical processing used to extract the refractory  grains from the 
bulk meteorite. Interstellar dust that entered into the solar nebula would 
have been reprocessed through many sublimation and recondensation events in 
the ISM, and would therefore have lost much of the chemical memory of its 
circumstellar birthsites.  

%
\begin{table}
\caption{Pre-solar Grains in the Orgueil Meteorite}
\begin{center}
\begin{tabular}{lcc}
  \hline
  \multicolumn{1}{c}{Composition} & 
  \multicolumn{1}{c}{Radius} & 
  \multicolumn{1}{c}{Abundance (\% by mass)} \\
  \hline
   Diamond     & $\sim 5$\AA   &  0.145              \\
   SiC         & 150\AA$-$5\um &  0.0014            \\
   Graphite    & 0.4$-$3.5\um  &  0.0006            \\
   Al$_2$O$_3$ & $\geq 0.5$\um &  $5 \times 10^{-5}$ \\
   TiC         & 35$-$100\AA   &  $\sim 10^{-9}$    \\
  \hline
\end{tabular}
\end{center}
\end{table}

In addition, to the meteoritic data there is now very strong evidence that 
the solar system is still collecting interstellar grains. The dust 
detectors on the  Ulysses spacecraft which flew by Jupiter in 1992 [27,28] 
detected sub-micron-sized particles, and the more recent radar studies of 
particles entering the Earth's atmosphere [29] detected small 
meteoroids. In both cases the particles had velocities in excess of 
the escape velocity for the solar system at the point of detection. The 
detected grains must therefore be of interstellar origin. The Ulysses 
particles  have a measured mass flux of $5 \times 10^{-21}$ g cm$^{-2}$ 
s$^{-1}$, roughly  consistent with the estimated flux of interstellar dust 
in the solar  vicinity [27]. The interstellar particles 
measured by the Ulysses fly-by of Jupiter have an average radius of 0.4\um, 
slightly larger than typical interstellar grains. The particles entering 
the Earth's atmosphere that were detected in the radar studies are 
significantly  larger (sizes $\sim $15--40\um) than those detected by the
Ulysses  experiments. However, grains smaller than those measured by Ulysses 
-- or typical interstellar grains, for that matter -- will be expelled from 
the inner solar system through the effects of radiation pressure and 
interaction with the solar wind carried magnetic field [28].

\section{Propagation Processes}

Grains formed in circumstellar shells propagate away from the star 
through the relatively gentle outward force of radiation pressure. 
In this process the grains couple with the gas through gas--grain 
collisions and drive mass loss. The radiation pressure force is 
counteracted by gas--drag. The net effect of these two forces determines 
the terminal velocity of the grains, and also the rate of grain growth in 
the densest parts of the circumstellar shell [30,31]. 
The outward radiation pressure force on the grains  is given by,
\begin{equation}
F_{\rm pr} = \pi a^2 \langle Q_{\rm pr} \rangle 
\left( \frac{L_{\star}}{4\pi r^2 c} \right), 
\end{equation}
where $a$ is the grain radius, $ \langle Q_{\rm pr} \rangle$ is the radiation 
pressure efficiency factor, $L_{\star}$ is the stellar luminosity, $r$ is 
the distance from the star, and $c$ is the velocity of light. The radiation 
pressure efficiency factor is determined by the absorption and scattering 
efficiency factors, $Q_{abs}$ and $Q_{sca}$, respectively, and 
a scattering asymmetry factor $g$, through the following expression,  
\begin{equation}
\langle Q_{\rm pr} \rangle = Q_{\rm abs} + ( 1-g ) Q_{\rm sca}.
\end{equation}
The parameter $g$ is a measure of the forward or backward-scattering 
properties of the grains and therefore, in part, determines the effectiveness 
of the coupling between the stellar radiation scattered by the grains and 
their outward motion. The gas drag force is approximately given by
\begin{equation}
F_{\rm d} \, =\, \pi a^2 \rho_{\rm gas} v_{\rm d}^2,
\end{equation}
where $\rho_{\rm gas}$ is the gas density, and $v_{\rm d}$ is the dust drift 
velocity with respect to the gas. The stellar mass loss rate is given by 
\begin{equation}
\dot M \, =\,  4 \pi \rho_{\rm gas} r^2 v
\end{equation}
where $v$ is terminal velocity of the gas, and we can then write the dust drift 
velocity (\cf [32]) as , 
\begin{equation}
v_{\rm d}\, \simeq \, \left( \frac{\langle Q_{\rm pr} \rangle L_{\star} v}
  {\dot M c}\right)^{1/2}\,
\simeq  1.9\, \left(\frac{a}{500{\rm\AA}}\right)^{1/2}\;\;\; 
  {\rm km\ s^{-1}},
\end{equation}
In Equation (5) we have adopted typical cool stellar radiation field 
parameters ($L_{\star}$ = 5000 L$_{\odot}$, $\dot M = 10^{-5}$
M$_{\odot}$ yr$^{-1}$, and $v$ = 10\kms), and a radiation pressure efficiency 
appropriate for silicates  ($\langle Q_{\rm pr} \rangle \simeq 0.02a/500$\AA).  
Carbonaceous particles have slightly higher efficiencies ($\langle Q_{\rm pr} 
\rangle \simeq 0.08a/500$\AA) and hence slightly higher drift velocities. 
From Equation (5) we see that the drift velocity increases with decreasing 
mass loss rate (\ie, density at the condensation radius) and for a mass 
loss rate of $10^{-8}$ M$_{\odot}$ yr$^{-1}$ reaches a limiting value of $\simeq$ 
40\kms\,for 500\AA\ silicate grains. At high drift velocities there is 
reduced dust--gas coupling and for lower mass loss rates there is not enough 
momentum flux in the dust to drive the gas in the form of a wind [32].  
In terms of the gas and dust balance of 
the ISM, the relatively few sources with the highest mass loss rates 
($>10^{-5}$ M$_{\odot}$ yr$^{-1}$) can dominate dust input to the ISM [33]. 
Typical drift velocities for these types of  envelopes are of order 1\kms, 
which is less than the threshold velocity for the sputtering of interstellar 
dust materials (typically 30\kms\,[34]). Sputtering 
destruction of the grains by impacting gas atoms in these environments 
is therefore unimportant. However, the lower threshold velocities 
($\simeq 2$\kms) for shattering in grain--grain collisions can lead to the 
fragmentation of  graphitic or silicate grains, but will be unimportant for 
the more resistant SiC or diamond grains [35]. 

Further processing can occur in the shock front ($v_{\rm s}\simeq$ 5--20\kms) 
where the ejecta merge with the ISM. Neglecting shattering, the processing of  
dust in the transition region between the stellar wind and the ISM has been 
considered [36], and it has been concluded that 
there is little dust destruction in these  environments. Our studies [35] 
show that, while shattering has a larger effect, the overall destruction and 
disruption is only minor in such low velocity shocks. Thus, it seems that 
stellar winds generally provide a gentle mechanism for dust propagation into 
the ISM.

Supernova shock waves are very efficient at propagating grains through
the ISM because they sweep up interstellar gas and dust, and accelerate 
the material to velocities of order tens to hundreds of kilometers per 
second. These shocks provide the kinetic energy that maintain the turbulent
motions in the ISM, and are thus responsible for the turbulent
diffusion of dust in the ISM.  Most of the mass of dust is in cold clouds,
both atomic and molecular; the shocks in these clouds are typically
of order 10\kms, fast enough to shatter about 1--10\% \ of the largest
grains but not to destroy them.  The shocks in the warm, intercloud medium 
are faster, and the large differential gas--grain and grain--grain velocities 
that are generated behind the shock front can lead to grain destruction and 
reprocessing. Thus, shock waves are essential in causing the dust to diffuse 
through the ISM, but not all dust survives this process. The destructive and 
disruptive effects of shock waves on dust  are considered in detail in the 
next section. In general the effects  of stellar jets on grains will be 
similar to the shocks considered below. However, the densities of the 
interaction regions will be higher than shocks in the warm medium, and the 
mass of the ISM affected by jets will be smaller than that affected by 
supernovae.

\section{Dust Survival in Shock Waves}

In shock waves, energetic grain--grain collisions and collisions between 
gas atoms/ions and dust grains lead to the loss of grain mass to the gas, 
and to changes in the grain size distribution. In Figure 2, we show the 
shock structure and postshock grain velocities for a 100\kms\,shock 
traversing the warm intercloud phase of the ISM [22,37]. The charged grains 
in the postshock gas undergo betatron acceleration as they gyrate around the 
magnetic field lines. As the postshock gas cools and compresses, the magnetic 
field increases, which in turn leads to the increased gyration speeds of the 
grains through the gas.  In Figure 2, two stages of betatron acceleration 
can be seen, each associated with increasing gas density. Betatron 
acceleration is opposed by collisional and plasma drag forces which try to 
bring the grains to rest with respect to the gas. At the higher grain 
velocities in Figure 2 the dominant drag force is that due to collisions 
with the gas, and varies as ($a\rho$)$^{-1}$, where $\rho$ is the grain 
material density. Thus, small and low-density grains are least affected by 
betatron acceleration and will be most resistant to destruction in shocks.

High energy collisions between gas atoms/ions and grains result in 
sputtering of the grain surfaces, \ie, the removal of grain surface species 
by sufficiently energetic collisions of atoms and ions. The sputtering may 
be thermal, due the high random kinetic velocities of the atoms and ions 
in the hot postshock gas, or inertial, due to the relative velocity of the 
betatron-accelerated grains with respect to the gas. The latter sputtering 
process is often referred to as non-thermal sputtering, but in this work we 
will use the more descriptive term of inertial sputtering.

Differential grain--grain velocities arise from the $(a\rho)^{-1}$ size and 
grain density dependence of the collisional drag of the gas atoms/ions on 
the grains.  Also, the grains are gyrating around the magnetic field lines 
and even without these drag effects the grains would have differential 
velocities.  In general, for the regions of a shock where grain 
destruction occurs, the velocity differential between grains increases 
with the difference in their radii (see Figure 2). Grain--grain collisions 
can lead both to vaporization, the transfer of grain mass to the gas as 
atoms/ions, and to shattering, the break-up of the colliding grains into 
smaller but distinct sub--grains (fragments). The velocity thresholds for 
vaporization and shattering are typically of order 20\kms\,and 2\kms, 
respectively [34,35]. The effects of shattering dominate over those of 
vaporization, and can lead to major redistributions of the grain mass [35,38].


\begin{figure}
 {\epsfxsize=\textwidth \epsffile{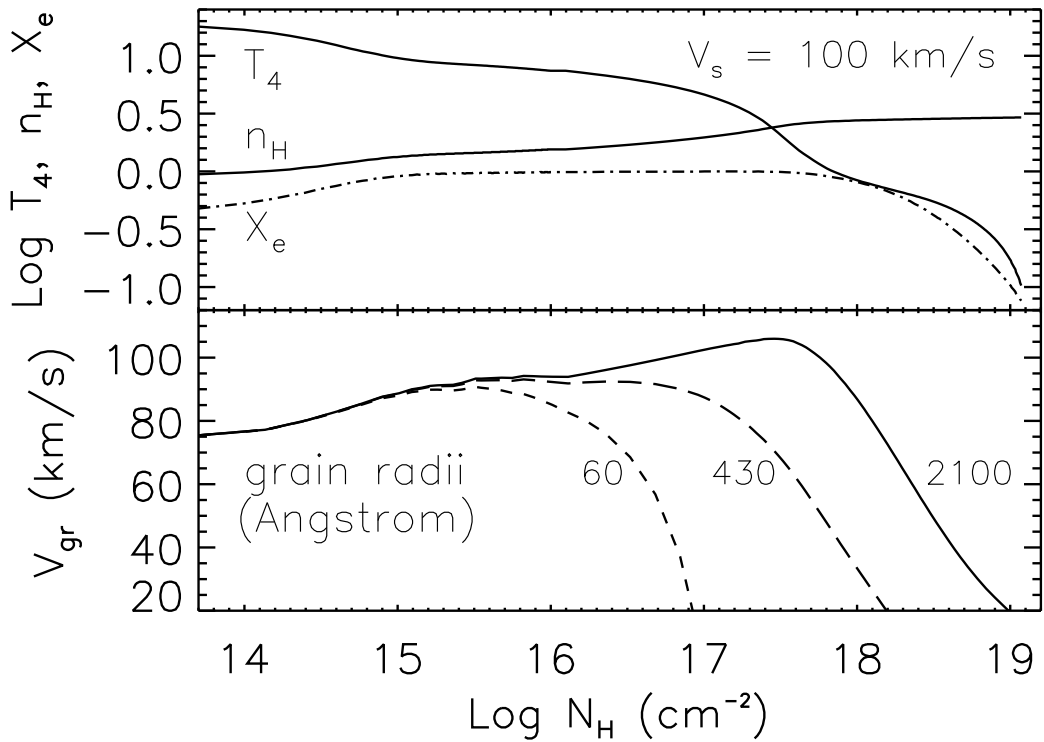}}
 \caption{ 100\kms\,shock profile (temperature, $T_4 = T_{\rm k}/10^4$K, 
 density, $n_{\rm H}$ (cm$^{-3}$), and electron relative abundance, 
 $X_{\rm e}$) as a function of the shocked column density $N_{\rm H}$. 
 Also shown in the lower plot are the graphite grains velocities as a 
 function of the shocked column density for three grain radii.}
\end{figure}

In keeping with previous work [35], we use the term ``destruction" to refer 
to the sputtering and vaporization processes that result in the transfer of 
grain mass to the gas; we use the term ``disruption"  for the shattering 
process, in which the total grain mass is preserved but redistributed over 
the grain size distribution. Disruption is any process where a grain is broken 
down into smaller fragments, and thus includes the fragmentation of ``solid'' 
particles and the disaggregation or breakup of porous, aggregated or cluster 
particles.

There have been many theoretical studies of the effects of supernova shock 
waves on interstellar dust [18,19,39--43]. We have undertaken a theoretical 
study of the effects of steady state, radiative, J shocks on the dust in the 
warm ISM [22,34,35,37]. As discussed above, it is in the warm medium 
where interstellar grain destruction by supernova shocks predominates.
This study uses up-to-date algorithms for the dust processing in shocks 
[34], and for the first time includes the effects of shattering in grain--grain 
collisions in shock waves [35]. In the calculations we assumed a preshock 
density of $n_{\rm H} = 0.25$ \nod, a temperature of $T_{\rm k} = 8000$ K, a 
component of the interstellar magnetic field normal to the shock front 
of $B_0 = 3 \, \mu$G, and an initial MRN size distributions of graphite 
and silicate grains. We modeled the effects of thermal and inertial 
sputtering of grains, and vaporization and shattering in grain--grain 
collisions in shocks of velocities $v_{\rm s} =$ 50, 100, 150, and 200\kms, and a 
range of preshock densities.  In Figure 3 we present the processing of 
graphite grains as a function of grain radius for a 100\kms\,shock, and in 
Figure 4 we show the preshock and postshock grain size distributions for 
shocks of velocity 50, 100, and 200\kms. These figures clearly show the 
dominant effects of grain shattering in redistributing the grain mass 
from large grains ($a > 500$\AA) into smaller fragments, and also the 
effects of inertial sputtering in grain destruction. Vaporization makes a 
relatively minor contribution to grain destruction.


\begin{figure}
 {\epsfxsize=\textwidth \epsffile{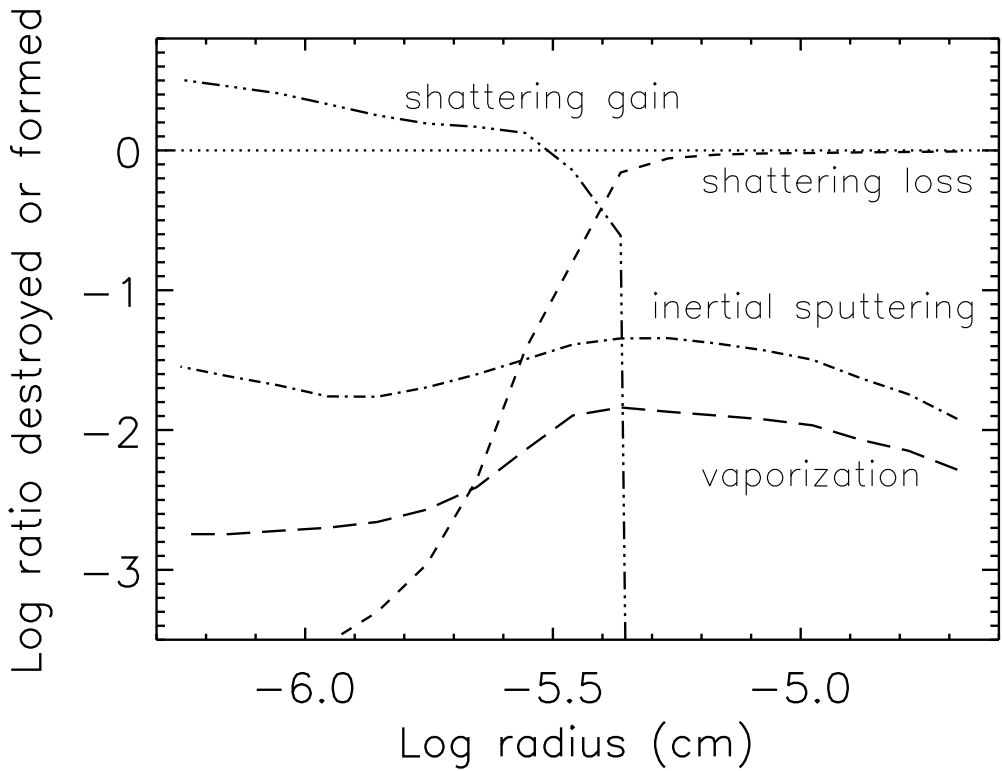}}
 \caption{Postshock graphite grain destruction, disruption and formation
 as a function of radius, given as the ratio of initial to final mass at
 a given radius, for vaporization ({\it long-dashed}), inertial
 sputtering ({\it dashed-dotted}), shattering mass loss ({\it short-dashed}),
 and shattering mass gain ({\it dashed-triple dotted}) in a 100\kms\,shock.}
\end{figure}

In Figure 5 we show the probability for interstellar grains to survive 
the passage of a supernova shock wave. This has been calculated as 
$a_{\rm half}$, the grain radius for which half of the initial grains 
survive. For grains with radii larger than $a_{\rm half}$ less than half 
of the original grains survive the passage of the shock, and for grains 
smaller than $a_{\rm half}$ more than half of the grains survive. From 
this one can see that large grains ($a > 1000$\AA) have a relatively 
high probability of surviving the passage of a 50\kms\,shock, 
but this probability decreases rapidly for higher shock velocities. 

We can calculate the time scale $t_{\rm SNR}$ for supernova shock waves to 
destroy interstellar dust (\ie, to return the grain mass to the gas as 
atoms) in all phases of the ISM [21]: 
\begin{equation}
t_{\rm SNR} = \frac{ 9.7 \times 10^7 }
{ \int \epsilon(v_{\rm s7})/v_{\rm s7}^{3}dv_{\rm s7} }~~~~~ {\rm yr} ,
\end{equation}
where $v_{\rm s7}$ is the shock velocity (in units of 100\kms) , and 
$\epsilon ( v_{\rm s7} )$ is the efficiency of grain destruction for a shock 
of velocity $v_{\rm s7}$.


\begin{figure}
 {\epsfxsize=\textwidth \epsffile{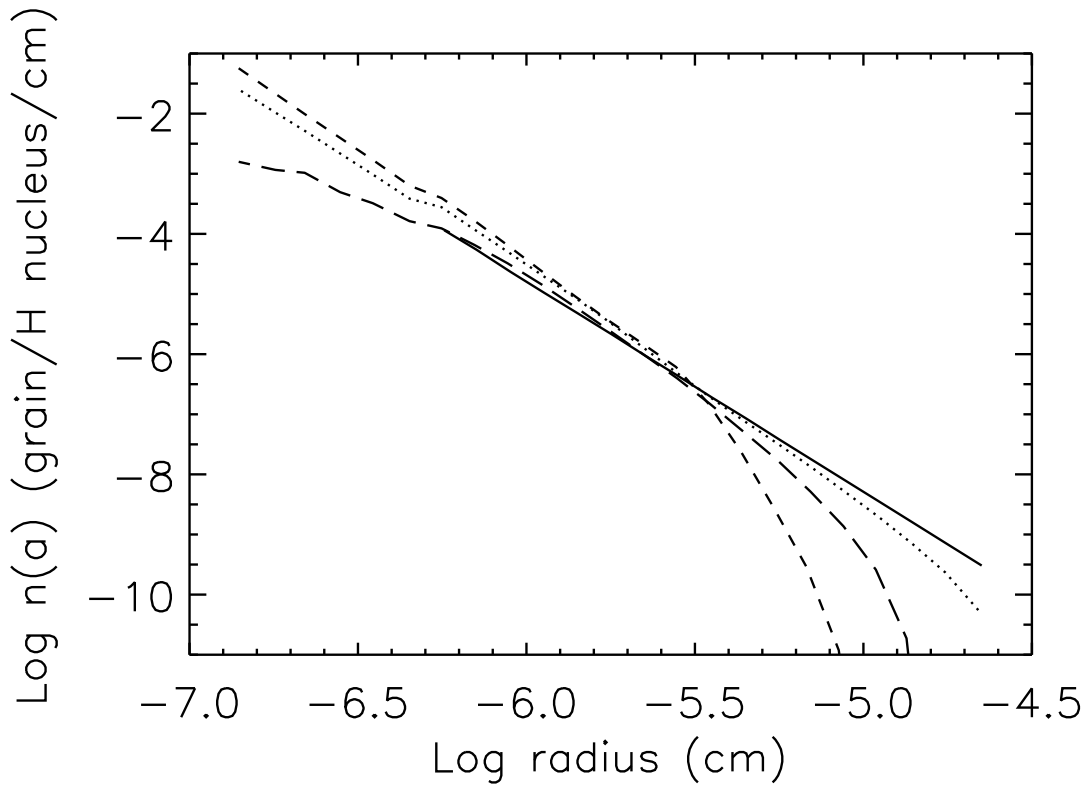}}
 \caption{Graphite grain initial MRN size
 distribution ({\it solid}), and postshock size distributions
 for shock velocities of 50\kms\,({\it dotted}),
 100\kms\,({\it short-dashed}), and 200\kms\,({\it long-dashed}).
 The preshock size distribution ranges from 50\AA\, to 2500\AA;
 $-6.3$ to $-4.6$ in Log radius (cm).}
\end{figure}

From calculated grain destruction data one can derive analytical
expressions for $\epsilon(v_{\rm s7})$ for the graphite and silicate MRN grain 
populations [22,35]. The time scales to return the entire grain mass to the 
gas as atoms are $t_{\rm SNR}=6.3 \times 10^8$ yr and $3.7 \times 10^8$ yr for 
graphite and silicate MRN size distributions, respectively. The destruction
of small grains ($<500$\AA) is dominated by high velocity ($>250$\kms)
shocks. Thermal sputtering in such a shock removes a layer of equal 
thickness, $\Delta a$, from all grains. For a 300\kms\,shock $\Delta a$ is 
of order 300\AA\ [20]. Thus, while high velocity shocks are very 
infrequent (once every $\simeq 6 \times 10^8$ yr), they essentially 
completely destroy all grains with sizes less than $\Delta a$. For large 
grains, betatron acceleration is important and shocks at all velocities
contribute about equally to the total destruction rate. In this case, the 
small fraction destroyed by a single low velocity shock, as compared to a 
high velocity shock, is compensated for by the much higher frequency of 
low velocity shocks. The average time between shocks of velocity $v_{\rm s7}$ 
is given by $10^8 v_{\rm s7}^2$ years [20]. In our earlier studies that 
did not include the effects of shattering large grain destruction was 
dominated by vaporization in grain--grain collisions in low velocity 
shocks ($v_{\rm s} = 50$\kms), and by sputtering in higher velocity shocks. 
In the latest models large grains are shattered in grain--grain collisions 
in the surface layers of shocks (50\kms\,$\leq v_{\rm s} \leq$ 200\kms). 
In high velocity shocks ($v_{\rm s} \geq$ 200\kms) the increased surface area 
to mass ratio, due to the fragmentation of the large grains, leads to 
much greater destruction by thermal sputtering in the hot postshock gas
[35], compared to the earlier models without shattering [22].


\begin{figure}
 {\epsfxsize=\textwidth \epsffile{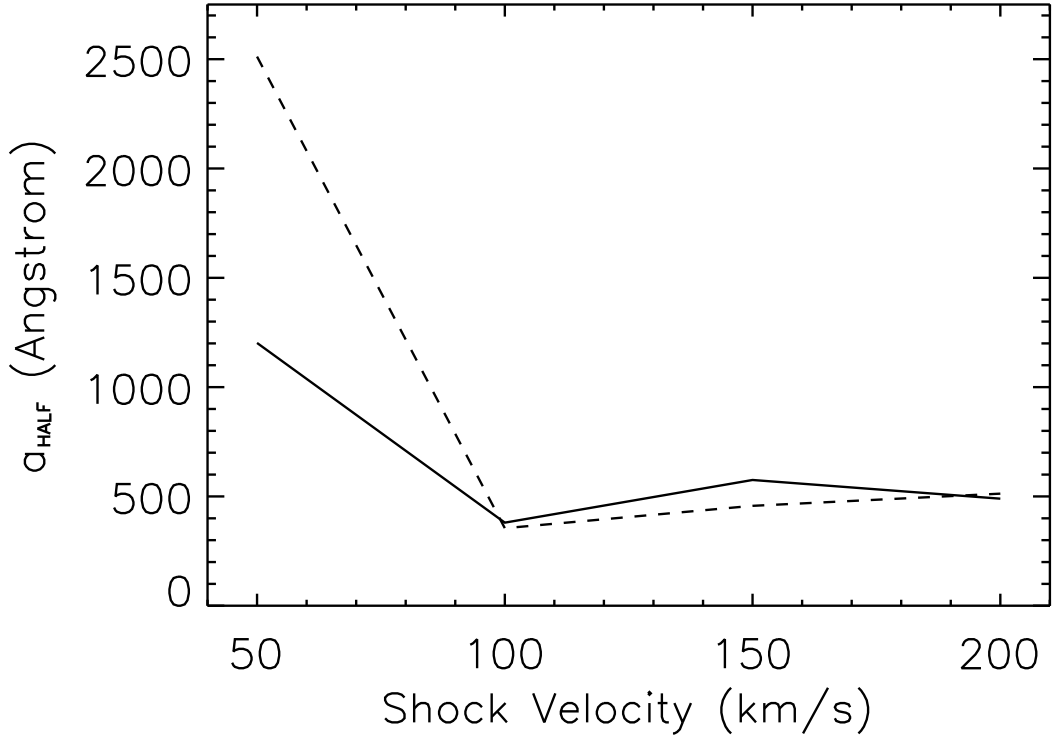}}
 \caption{Grain radius for which 50\% of the initial grains survive the passage
 of a single shock, $a_{\rm HALF}$, plotted against the shock velocity, for
 graphite/amorphous carbon grains ({\it solid line}) and silicate grains
 ({\it dashed line}). Grains larger than $a_{\rm HALF}$ have a probability of
 less than half of surviving the passage of a single shock.}
\end{figure}

The inclusion of shattering in the shock code increases the derived grain 
lifetimes slightly, compared to calculations without shattering, because 
the large grains are shattered into smaller particles that survive better 
in shocks.  However, these lifetimes still fall short of the dust lifetimes 
of $t_{\rm SNR} \geq 2 \times 10^{10}$ yr [22] required to preserve $\geq$ 90\% 
of the stellar ejected silicates in the ISM [5,44]. Therefore, the 
well-known conundrum remains [18,22,37,39--43]: 
Why is silicate dust, which may contain as much as 90\% of 
the available silicon, so abundant in the ISM? Moreover, a marginally 
increased grain lifetime is now accompanied by  the almost complete 
disruption of large interstellar grains (1000\AA\, $\leq a \leq$ 2500\AA).
Shattering in grain--grain collisions therefore adds a new conundrum: 
Why do visual extinction measurements show that most of the mass in 
interstellar dust is in large grains (radii $\geq$ 1000\AA)?

Shattering in grain--grain collisions leads to the almost complete elimination 
of large grains ($a \geq 1000$\AA) from the MRN dust population in a fast shock 
($v\sim 100$\kms). 
The disruption time scales, $t_{\rm dd}$ --- the 
time scale to disrupt all large grains in the ISM and to 
shatter them into sub-500\AA\, fragments in grain--grain collisions in shock 
waves --- are considerably shorter than the destruction time scales.  The 
derived grain lifetimes, $t_{\rm SNR}$, and large--grain disruption time scales, 
$t_{\rm dd}$, are summarized in Table 3. The disruption time scales are factors 
12 (graphite) and 6 (silicate) times smaller than the destruction lifetimes.
We conclude from this that large graphite and large silicate 
grains have short disruption time scales in the diffuse ISM.
This implies that large interstellar grains must reform in the 
ISM. This can occur by the condensation of volatile gas phase species onto 
grain surfaces (accretion), or by the coagulation of the grains into large 
porous structures. Accretion alone cannot cannot reform large grains because 
of the limited supply of condensible gas phase species. The accretion of icy 
mantles may, however, aid coagulation by increasing the probability of 
particle coalescence in grain--grain collisions [45]. The monomers or cores 
making up these composite aggregates are likely have radii less than about 
500\AA, as indicated by Figure 5.

%
\begin{table}
\caption{Interstellar Dust Lifetimes, $t_{\rm SNR}$, and Disruption 
 Time Scales, $t_{\rm dd}$ }
\begin{center}
\begin{tabular}{lcc}
  \multicolumn{1}{c}{ } & \multicolumn{2}{c}{$t_{\rm SNR}$ ($\times 10^8$ year)} \\
  \hline
  \multicolumn{1}{c}{Component} & 
  \multicolumn{1}{c}{Carbon} & 
  \multicolumn{1}{c}{Silicate} \\
  \hline
   MRN            & 6.3        & 3.7        \\
  \hline \\[0.1cm]
  \multicolumn{1}{c}{ } & \multicolumn{2}{c}{$t_{\rm dd}$ ($\times 10^8$ year)} \\
  \hline
   $\geq 1000$\AA & $\leq 0.5$ & $\leq 0.7$ \\
  \hline \\[0.1cm]
  \multicolumn{3}{l}{MRN \hspace*{0.4cm} = Mathis, Rumpl \& Norsieck (1977)} \\
  \multicolumn{3}{l}{\hspace*{1.9cm} size distribution, $dn(a) \propto a^{-3.5} 
   da$,} \\ 
   \multicolumn{3}{l}{\hspace*{1.9cm} with 50\AA\, $\leq a \leq 2500$\AA} \\
  \multicolumn{3}{l}{$\geq 1000$\AA\, = all grains in the MRN size} \\
  \multicolumn{3}{l}{\hspace*{1.9cm} distribution with radii $\geq 1000$\AA} \\
\end{tabular}
\end{center}
\end{table}

\section{Mitigating Effects on Dust Processing in the ISM}

All theoretical studies of dust processing in the ISM contain certain
assumptions about the values of physical parameters that are not
well-determined observationally or experimentally. It is therefore
worthwhile to consider and discuss some of these uncertainties, and
to see how they may effect the derived time scales for dust processing
in the ISM. These regions of uncertainty fall into three broad categories,
namely: the structure of the ISM, the structure of supernova remnant shocks,
and the microphysical properties of the dust. In this section we discuss
some of the uncertainties in these parameters and their effects on the dust
processing time scales.

Firstly, with regard to the structure of the ISM, the volume filling factors
for the different components, and the cycling time scales between the
components are not well-determined observationally. For example, in our
calculations [1,22,35] we have assumed an idealized multiphase model of the
ISM [17], and the results of these calculations depend on the parameters of
the ISM adopted in the model. Most of the dust destruction occurs in the
warm intercloud gas, but the amount of dust destruction is not very
sensitive to the model of the ISM, being about the same for both two-phase
and three-phase models [21].  The rate of dust destruction would be reduced
if there were substantially less dust in the warm gas than in the cold gas,
but this is not the case [46]. Although the gaseous abundances of refractory
elements are significantly higher in the warm gas than in the cold gas, most
of the refractory material remains tied up in dust in both phases. The
calculations of the dust processing time scale linearly with the adopted
supernova rate. Some supernovae occur in stellar associations; since later
supernovae occur within the bubbles created by earlier supernovae in these
associations, such supernova will be less effective in destroying dust. This 
effect was allowed for in our estimate of the dust destruction time, which is 
based on [21]: we have used an effective supernova rate in the Galaxy of 1 per 
125 years, substantially less than the actual rate of about 1 per 40 yr 
(for H$_0=75$\kms\,Mpc$^{-1}$ [47]).

Secondly, dust destruction is directly affected by the structure of supernova
remnant shock waves.  For shock velocities greater than about 200\kms\,the
shock is non-radiative and the amount of dust destroyed depends upon the
long-term evolution of the remnant.  Thermal conduction and cloud evaporation
both reduce the temperature of the shocked gas and thereby reduce dust
destruction by thermal sputtering, the dominant dust destruction process in
such shocks. Slower shocks are radiative, and the amount of dust destroyed
depends upon the strength of the magnetic field [37]. If the Galactic magnetic 
field is about 5\,$\mu$G instead of 3\,$\mu$G [48], then the dust destruction 
in radiative shocks will be reduced by a factor of about 2 [22]; the overall 
reduction is then substantially less than the factor of 8 that has been 
suggested [49]. The calculations undertaken to date assume that the supernova 
remnant expands into a uniform medium or a uniformly cloudy medium. Massive 
stars will create low density cavities surrounded by shells of gas [50], and 
it is not known yet how this will affect the amount of grain destruction.

Thirdly, the microphysical properties of the dust will directly determine 
its processing time scales. Recent theoretical studies of the sputtering of
astrophysically relevant materials suggest much lower yields near threshold
[51] than our semi-empirical sputtering formalism predicts [34]. For
carbonaceous grains, much of the discrepancy reflects a difference in the
adopted binding energy (7.4\,eV appropriate for graphite [51] versus 4\,eV
appropriate for amorphous carbon [52]). It is well established experimentally
that graphite will amorphitize under the effects of ion-bombardment and, thus, 
it has been argued [34] that the lower binding energy is the correct one to 
use, even if the preshock dust is assumed to be graphite. For silicates, the 
origin of the discrepancy is not well understood but it should be noted that 
the theoretical method may be suspect near threshold [53]. In the end, these 
differences in the adopted binding energies or sputtering yields near 
threshold do not influence the derived dust life time very much. While the 
exact threshold influences the minor fraction of dust returned to the gas at 
low shock velocities, the dust lifetime measures the contribution from all 
shocks where the grain velocities may substantially exceed this threshold, 
and in this case the influence of the threshold is minimal. In fact, while 
silicate and carbon grains have binding energies of 5.7\,eV and 4\,eV in our 
calculations, the calculated lifetime of silicate grains is slightly shorter 
than for carbon ($4\times 10^8$ yr versus $6\times 10^8$ yr), reflecting the 
difference in specific density and hence stopping length behind the shock [22]. 
In principle, the structure of interstellar grains (\ie, porosity) has more  
consequences for the calculated lifetimes. The mass to cross-section 
ratio affects the grain stopping length, and hence its destruction, because 
the drag on the grains varies as $(a\rho)^{-1}$. Moreover, porous aggregate 
particles may be disrupted (shattered) early-on in the shock and, therefore, 
are not betatron accelerated. This will limit the destruction of such grains 
in radiative shocks ($v_{\rm s} < 200$\kms). However, by the same token, the 
much larger surface area offered by the disrupted aggregates will lead to 
greater thermal sputtering in non-radiative shocks ($v_{\rm s} > 200$\kms). 
While we have yet to carry out a full calculation of these effects, we 
anticipate that the difference in the derived dust lifetimes will be less 
than a factor two.

Clearly, there are many variables in the determination of dust processing time
scales in the ISM that need to be much better determined, \eg, the structure
and evolution of supernova remnants and the physical processing of porous dust
particles in shock waves. Therefore, it is in these directions that future
studies of the ISM and dust processing within it could most usefully be
directed. Nevertheless, as emphasized in this review, dust lifetimes in the
ISM are predicted to be $\simeq 5\times 10^8$ yr, much shorter than the
stardust injection time scale of $\simeq 3 \times 10^9$ yr [1]. As a direct
consequence, dust reformation in the ISM is a necessary condition for the
evolution of interstellar dust.

\section{Discussion}

Figure 3 shows that graphite (or amorphous carbon) grains, lose 2--5\% of 
their mass, or a fraction of their radius $\Delta a/a \simeq$ 0.7--1.7\% , 
due to inertial sputtering in a 100\kms\,shock.  This primarily occurs 
through the process of sputtering by gas atoms and ions  during slowing 
down with respect to the gas. The mass loss due to inertial sputtering is 
relatively insensitive to the grain size. Essentially, grains collide with 
approximately their own mass of gas during the slowing down process. 
Betatron acceleration increases this mass loss somewhat for the larger 
grains for shock velocities, $v>50$\kms. For SiC grains the inertial 
sputtering yield is about 50\% higher than for graphite [34], and so 
$\Delta a/a$ will be of order 1.0--2.2\%. Diamond materials have similar 
sputtering yields to graphite, and so the  degree of inertial sputtering 
for the interstellar diamonds will be the same as for graphite, \ie, 
$\Delta a/a \simeq$ 0.7--1.7\%. 

For large grains shattering due to grain--grain collisions has a profound
influence. The largest grains considered in shock calculations are generally 
those defined by a MRN size distribution, \ie, of order a few  thousand \AA.
These grains are rapidly disrupted into smaller grain fragments (see 
\S 4), and so are predicted to be rapidly removed from the interstellar size 
distribution on time scales of order a few tens of millions of years by 
shocks in the warm ISM [35]. They are primarily disrupted by catastrophic 
grain--grain collisions; collisions in which more than half of the larger 
(target) grain mass is shattered into sub-300\AA\, fragments. However,  
interstellar grains with sizes $\gg$ 1\um\ will fare somewhat better 
in interstellar shocks because of their large mass and the low abundance 
of collision partners large enough to cause their catastrophic 
fragmentation in grain--grain collisions. In general, the main agent for the 
disruption of a moving target grain is a collision with a grain which is 
just large enough for the kinetic energy of the impact to catastrophically 
shatter the whole grain. For a 5\um\ SiC grain at 100\kms, 
this grain limiting size is $\simeq$ 0.8\um\ as compared to  a limiting
size of 500\AA\ for a typical interstellar grain of 2500\AA . Because the
interstellar grain size distribution steepens considerably above 
$a = 3000$\AA , a 5\um\ grain will have a very low probability for such a 
catastrophic collision to occur. Such large grains will still being cratered 
by impacts with the smallest grains and lose $\simeq$ 5\% of their mass. Thus, 
large grains travel relatively unscathed through interstellar shocks. In 
contrast, because of their lower intrinsic strength, large ($<10$\um) 
graphite or silicate grains will be disrupted in shocks. 

It is an inescapable conclusion of the analysis presented here that 
presolar grains should retain some memory of their sojourn through the 
ISM, \ie, traces of exposure to sputtering in the gas phase and 
surface pitting due to the sub-catastrophic (or cratering) impact of 
small grains in shock waves. The inferred ages for the recovered SiC 
grains are of the order of hundreds of millions of years [54] 
much larger than the typical interval between supernovae 
shock waves ($\sim$ few $\times 10^7$ years). The extracted presolar 
grains would therefore be expected to show some traces of their passage 
through the ISM. However, it appears that the extracted grain surfaces 
seem to be relatively clean, and that many of the particles have 
crystallographic faces (Bernatowicz these proceedings), which is clearly 
in contrast to the expectations from the discussion above.  It should, 
nevertheless,  be pointed out that in extracting the presolar grains from 
the meteoritic material they undergo very extreme oxidation processes, 
which may in part modify the original surfaces of the particles. Perhaps 
the surfaces seen in the laboratory are not the same faces that were 
exposed to the ISM?  It is also possible that in the general ISM the grains 
are mantled with some other material. This could perhaps explain why the 
SiC absorption feature at 11.3\um\ has only been seen in circumstellar 
environments; outside these regions the presence of absorbing mantles of 
carbon could suppress the feature [55]. The most likely 
solution, particularly for the large graphite grains, is that the 
recovered grains are those that  have never been exposed to interstellar
shock waves; i.e., the grains that survive are the (few) grains which 
never saw a strong shock in the first place. This is also true for the 
small grains (\ie, the diamonds) which are relatively unaffected by 
low velocity shocks ($<$ 250\kms) but are completely destroyed by the 
first high velocity shock they encounter. With a fast shock time scale of 
$3 \times 10^8$ yr and a star-ISM cycle time of $3 \times 10^9$ yr, an 
average sample of the ISM would contain $\simeq 10$\% of material which 
never saw such a fast shock.

\section{Conclusions}

Stellar winds are the initial injectors of circumstellar dust from
its site of formation into the ambient ISM. The action
of the stellar radiation on the dust, and its collisions with the gas 
are relatively benign resulting in very little processing during this 
stage in its evolution. However, stellar winds are probably not able 
to propagate the dust over large interstellar distances. For dust 
distribution over large interstellar scales we require that the effects 
of supernovae play a major role. Supernova shock waves can drive large 
masses of gas and dust through the ISM, but they also heavily process 
or even destroy the bulk of the entrapped dust.

Within the context of the sizes and composition of the extracted 
meteoritic presolar grains we can now draw some interesting conclusions. 
These grains must have undergone some processing in supernova shock waves 
in the warm ISM, the phase in which dominant grain destruction occurs. 
The results of model calculations show that sub-100\AA\, refractory grains 
such as the diamonds recovered from meteorites can survive passage 
through shock waves ($v_{\rm s} < 200$\kms) relatively unharmed. They 
merely lose of order 1--2\% of their outer layers, which for a 100\AA\,
radius grain would be less than one atomic layer, and therefore of minimal 
consequence. On the other hand thermal sputtering in fast shocks does destroy 
small grains efficiently. However, $\simeq 10$\% of the interstellar grains 
are incorporated into stars and planetary systems before they encounter 
such a fast shock. For micron-sized SiC grains the dominant surface 
processing will be due to inertial sputtering, because of their great 
resistance to cratering in collisions with small grains. However, for 
micron-sized graphite grains the effects of cratering should dominate. 
Perhaps these differences are reflected in the nature of the meteoritic 
presolar grain surfaces?  However, the indications from extensive sample 
analyses are that the surfaces of the recovered presolar grains are 
unprocessed. Does this result really imply that the grains have not been 
processed in the ISM, or that the surfaces of the analyzed grains are not 
the faces that were exposed in the ISM? Perhaps modification in the 
laboratory, during the rigorous extraction processes, has removed the 
original surfaces? What is now seen could be some underlying layer that 
was protected by overlying layers, or by mantles of material deposited in 
the ISM. In order to address these questions concerning the surface 
processing of dust in the ISM it would be useful to perform some control 
experiments on laboratory-made micron-sized particles. It is possible that 
during the extraction of interstellar grains from meteorites some surface 
alteration may occur. The same oxidative extraction processes used to 
recover interstellar meteoritic grains should be applied to 
well-characterized SiC and graphite particles, in order to quantify the 
degree of surface alteration that occurs during the extraction process. 

The presolar grains extracted from meteorites indicate that these grains 
can survive for long time scales in the ISM, survive the 
process of incorporation into the solar nebula, and also survive eventual 
capture into meteoritic bodies in the solar system. However, it must 
be remembered that the extracted presolar grains consist of highly  
refractory materials, and are thus a very selective sample of the dust 
in the general ISM. Also, the extracted dust may represent a small 
fraction of the interstellar stardust which did survive shock 
processing. The bulk of the interstellar grain materials, and those for 
which we have observational evidence, \eg, amorphous silicates, and by 
inference also amorphous carbon, have not yet been extracted from the 
meteorites and analyzed. This is probably no easy task, given their less 
resilient nature. Capturing interstellar grains in space and returning 
them to Earth for analysis is required if we are to obtain a more
representative sample.

\vspace*{0.5cm}
\noindent{\bf Acknowledgements.}
APJ wishes to thank the McDonnell Center for the Space Sciences and NASA
for the invitation to attend this conference.  Theoretical studies of
interstellar dust at NASA Ames are supported under RTOP 399-20-10-13 from the
astrophysics theory program. The research of CFM is supported in part by NSF
grant AST95-30480.


\newpage
\noindent {\large \bf References}
\vspace*{0.25cm} \noindent 
\newline 1. Jones, A.P., \& Tielens, A.G.G.M. 1994, in The Cold Universe,
           XIIIth Moriond \newline \hspace*{0.5cm}Astrophysics 
           Meeting, ed. T. Montmerle, C.J. Lada,
           I.F. Mirabel, \& J. Tran 
           \newline \hspace*{0.5cm}Thanh Van (Gif-sur Yvette: Editions 
           Frontieres), 35
\newline 2. Jones, A.P. 1997, in From Stardust to Planetesimals, 
           ed. Y.J. Pendleton, \&  \newline \hspace*{0.5cm}A.G.G.M. Tielens, 
           Astronomical Society of the Pacific, Conference Series, in press
\newline 3. Cherchneff, I., \& Tielens, A.G.G.M. 1994, in Circumstellar Media 
           in the Late \hspace*{0.5cm}Stages of Stellar Evolution, ed. R.E.S. 
           Clegg and W.P.S. Meikle, (Cambridge \hspace*{0.5cm}University 
           Press), p232
\newline 4. Moore, M.H., Tanabe, T., \& Nuth, J.A. 1991, \apj, 373, L31
\newline 5. Mathis, J.S. 1990, \araa, 28, 37
\newline 6. Mathis, J.S., Rumpl, W., \& Nordsieck, K.H. 1977, \apj, 217, 105 (MRN) 
\newline 7. Tielens, A.G.G.M. 1990, in Carbon in the Galaxy, ed. J.C. Tarter,
           S. Chang, \& \hspace*{0.5cm}D. DeFrees (NASA Conf. Publ. 3061), 59
\newline 8. Kim, S.-H., Martin, P.G., \& Hendry, P.D. 1994, \apj, 422, 164
\newline 9. Whittet, D.C.B. 1992, Dust in the Galactic Environment, Institute
           of Physics \newline \hspace*{0.5cm}Publishing, Bristol.
\newline 10. Millar, T.J., \& Williams, D.A. (eds.) 1993, Dust and Chemistry in
           Astrophysics, \newline \hspace*{0.6cm}Institute of Physics Publishing, 
           Bristol
\newline 11. Evans, A. 1994, The Dusty Universe, John Wiley, Chichester 
\newline 12. Cutri, R.M., \& Latter, W.B. (eds.) 1994, The First Symposium 
           on the Infrared \newline \hspace*{0.6cm}Cirrus and 
           Diffuse Interstellar Clouds, San Francisco,
           A.S.P. Conference Series \newline \hspace*{0.6cm}58
\newline 13. Nenner, I. (ed.) 1994, Molecules and Grains in Space, AIP Conf. Proc. 
           312, \newline \hspace*{0.6cm}AIP Press, New York
\newline 14. Greenberg, J.M. (ed.) 1995, The Cosmic Dust Connection, Kluwer, 
           Dordrecht
\newline 15. Dorschner, J., \& Henning, T. 1995, \aap\ Reviews, 6, 271
\newline 16. Pendleton, Y.J., \& Tielens, A.G.G.M. (eds.) 1997, From Stardust to 
            Planetesi- \newline \hspace*{0.6cm}mals Astronomical Society 
            of the Pacific, Conference Series
\newline 17. McKee, C.F., \& Ostriker, J.P. 1977, \apj, 218, 148 
\newline 18. Draine, B.T., \& Salpeter, E.E. 1979, \apj, 231, 438
\newline 19. Dwek, E., \& Scalo, J.M. 1979, \apj, 233, L81
\newline 20. Seab, C.G. 1987, in Interstellar Processes. ed. D.J. Hollenbach
           \& H.A. Thron- \newline \hspace*{0.6cm}son Jr., 
           (Dordrecht: Reidel), 491
\newline 21. McKee, C.F. 1989, in Interstellar Dust. ed. L.J. Allamandola \& 
            A.G.G.M. \newline \hspace*{0.6cm}Tielens, (Dordrecht, Kluwer), 431
\newline 22. Jones, A.P., Tielens, A.G.G.M., Hollenbach, D.J., \&
           McKee, C. F. 1994, \apj, \newline \hspace*{0.6cm}433, 797 
\newline 23. Bradley, J.P. 1994, Science, 265, 925 
\newline 24. Martin, P.G. 1995, \apjl, 445, L63
\newline 25. Anders, E., \& Zinner, E. 1993, Meteoritics, 28, 490
\newline 26. Draine, B.T. 1995, in The Physics of the Interstellar Medium 
           and Intergalactic  \newline \hspace*{0.6cm}Medium, A. Ferrara, 
           C.F. McKee, C. Heiles, \& P.R. Shapiro (eds.), Astronomical 
           \newline \hspace*{0.6cm}Society of the Pacific, 
           Conference Series, Vol. 80, 133 

\newpage

\noindent
\newline 27. Gr\"{u}n, E., Zook, H.A., Baguhl, M., Balogh, A., Bame, S.J., 
           Fechtig, H., Forsyth, \newline \hspace*{0.6cm}R., Hanner, M.S.,
           Horanyi, M., Kissel, J., Lindblad, B.-A., Linkert, D., Linkert, 
           \newline \hspace*{0.6cm}G., Mann, I., McDonnell, 
           J.A.M., Morfill, G.E., Phillips, J.L., Polanskey, C., 
           \newline \hspace*{0.6cm}Schwehm, G., Siddique, N., Staubach, P., 
           Svestka, J., \& Taylor, A. 1993, \newline \hspace*{0.6cm}Nature, 
           362, 428 
\newline 28. Gr\"{u}n, E., Gustafson, B., Mann, I., Baguhl, M., Morfill, G.E., 
           Staubach, P., \newline \hspace*{0.6cm}Taylor,  A., \& Zook, 
           H.A. 1994, \aap, 286, 915
\newline 29. Taylor, A., Baggaley, W.J., \& Steel, D.I. 1996, Nature, 380, 323
\newline 30. Tielens, A.G.G.M. 1983, \apj, 271, 702
\newline 31. Dominik, C., Gail, H.-P., \& Sedlmayr, E. 1989, \aap, 223, 227
\newline 32. Habing, H.J., Tignon, J., \& Tielens, A.G.G.M. 1994, \aap, 286, 523
\newline 33. Jura, M. 1987, in Interstellar Processes, ed. D.J. Hollenbach \&
           H.A. Thronson, \newline \hspace*{0.6cm}(Dordrecht; Reidel), 3
\newline 34. Tielens, A.G.G.M., McKee, C.F., Seab, C.G., \& Hollenbach, D.J.
           1994, \apj, \newline \hspace*{0.6cm}431, 321
\newline 35. Jones, A.P., Tielens, A.G.G.M., \& Hollenbach, D.J. 1996, 
           \apj, 469, 740
\newline 36. Woitke, P., Dominik, C., \& Sedlmayr, E. 1993, \aap, 274, 451 
\newline 37. McKee, C.F., Hollenbach, D.J., Seab, C.G., \& Tielens, A.G.G.M.
           1987, \apj, \newline \hspace*{0.6cm}318, 674
\newline 38. Borkowski, K.J., \& Dwek, E. 1995, \apj, 454, 254
\newline 39. Barlow, M.J. 1978, \mnras, 183, 367
\newline 40. Barlow, M.J. 1978, \mnras, 183, 397
\newline 41. Draine, B.T., \& Salpeter, E.E. 1979, \apj, 231, 77
\newline 42. Dwek, E., \& Scalo, J.M. 1980, \apj, 239, 193
\newline 43. Seab, C.G., \& Shull, J.M. 1983, \apj, 275, 652
\newline 44. Draine, B.T., \& Lee, H.K. 1984, \apj, 285, 89
\newline 45. Chokshi, A., Tielens, A.G.G.M., \& Hollenbach, D. 1993, \apj, 
           407, 806
\newline 46. Savage, B.D., \& Sembach, K.R. 1996, \apj, 470, 893
\newline 47. van den Bergh, S., \& McClure, R.D. 1994, ApJ 425, 205
\newline 48. Boulares, A., \& Cox, D.P. 1990, \apj, 365, 544
\newline 49. Crinklaw, G., Federman, S.R., \& Joseph, C.L. 1994, \apj, 424, 748
\newline 50. McKee, C.F., Van Buren, D., \& Lazareff, B. 1984, ApJ 278, L115
\newline 51. Flower, D.R., Pineau des For\^{e}ts, G., Field, D., \& May, P.W. 
           1996, \mnras, \newline \hspace*{0.6cm}280, 447
\newline 52. Benson, S.W. 1976, Thermochemical Kinetics (New York: Wiley)
\newline 53. Tielens, A.G.G.M. 1997, in Formation and Evolution of Solids in
           Space, ed. \newline \hspace*{0.6cm}J.M. Greenberg, in press
\newline 54. Lewis, R.S., Amari, S., \& Anders, E. 1990, Nature, 348, 293
\newline 55. Kozasa, T., Dorschner, J., Henning, T., \& Stognienko, R. 
           1996, \aap, 307, 551
\end{document}